\documentclass[twocolumn,showpacs,preprintnumbers]{revtex4}

\input{tcilatex}

\begin{document}

\bigskip

\title{Generalized Schwinger boson mean field theory for two-dimensional
SU(4) symmetric spin-orbital antiferromagnet}
\author{Guang-Ming Zhang}
\address{Center for Advanced Study, Tsinghua University, 
Beijing 100084, China}
\date{\today}

\begin{abstract}
A generalized Schwinger boson representation is proposed to describe the
two-dimensional SU(4) symmetric spin-orbital quantum antiferromagnet. A
uniform mean field solution gives rise to an antiferromagnetic long range
ordered state with a small staggered magnetization $m_{s}=0.07566$, and the
recent quantum Monte Carlo simulations on rather large lattice sizes have
provided strong evidence to support these mean field results.
\end{abstract}

\maketitle

There have been much interests in the properties of Mott insulators with
orbital degeneracy. In order to describe low-energy physics of an insulating
phase with one-electron per site with double orbital degeneracy in the limit
of the large Hubbard repulsion, an SU(4) symmetric model was proposed \cite%
{santoro} 
\[
H=\sum_{<i,j>}[\mathbf{S}_{i}\cdot \mathbf{S}_{j}+\mathbf{T}_{i}\cdot 
\mathbf{T}_{j}-4(\mathbf{S}_{i}\cdot \mathbf{S}_{j})(\mathbf{T}_{i}\cdot 
\mathbf{T}_{j})], 
\]%
where $\mathbf{S}_{i}$ and $\mathbf{T}_{i}$ denote the spin-1/2 and
orbital-1/2 operators at a lattice site $i$, respectively. The physical
condition to derive the above effective model is that among the possible
two-particle states obtained upon virtual hopping, the inter-orbital singlet
is the lowest energy state due to the dynamic Jahn-Teller effect.

Recently, by generalizing the SU(2) Schwinger boson representation \cite%
{aa1988}, we proposed an SU(4) spin-boson representation to denote\ both the
spin-1/2 and orbital-1/2 operators \textit{simultaneously}, and the model is
simplified to a quadratic form of a nearest neighboring SU(4) symmetric
valence bond (VB) pairing operator \cite{zhang-shen}. By a direct
Hartree-Fock decomposition and replacing the local constraint by its bulk
average, a uniform SU(4) mean field (MF) theory gives rise to a spin-orbital
liquid state with a finite gap in 1D, consistent with the exact matrix
product solution \cite{martins} and the numerical simulations \cite{santoro}%
. On a 2D square lattice, the spin, orbital, and spin-orbital tensor form
antiferromagnetic (AF) long range ordering state.

It is noted that total spin, orbital, and \textit{staggered} spin-orbital
operators $\sum_{j}S_{j}^{\alpha }$, $\sum_{j}T_{j}^{\alpha }$, $%
\sum_{j}2S_{j}^{\alpha }T_{j}^{\beta }e^{i\mathbf{Q\cdot R}_{j}}$ with $%
\mathbf{Q}$ the AF reciprocal vector, generate an SU(4) Lie algebra and
commutes with the Hamiltonian. In order to characterize both spin and
orbital degrees of freedom simultaneously, an SU(4) generalized Schwinger
boson representation is introduced as the generators $F_{\beta }^{\alpha
}(i)=b_{i,\alpha }^{\dagger }b_{i,\beta }$ with the expressions%
\begin{eqnarray}
&&S_{i}^{+}=b_{i,1}^{\dagger }b_{i,2}+b_{i,3}^{\dagger
}b_{i,4},S_{i}^{-}=b_{i,2}^{\dagger }b_{i,1}+b_{i,4}^{\dagger }b_{i,3}, 
\nonumber \\
&&S_{i}^{z}=(b_{i,1}^{\dagger }b_{i,1}-b_{i,2}^{\dagger
}b_{i,2}+b_{i,3}^{\dagger }b_{i,3}-b_{i,4}^{\dagger }b_{i,4})/2;  \nonumber
\\
&&T_{i}^{+}=b_{i,1}^{\dagger }b_{i,3}+b_{i,2}^{\dagger
}b_{i,4},T_{i}^{-}=b_{i,3}^{\dagger }b_{i,1}+b_{i,4}^{\dagger }b_{i,2}, 
\nonumber \\
&&T_{i}^{z}=(b_{i,1}^{\dagger }b_{i,1}+b_{i,2}^{\dagger
}b_{i,2}-b_{i,3}^{\dagger }b_{i,3}-b_{i,4}^{\dagger }b_{i,4})/2,
\end{eqnarray}%
with a local constraint $\sum_{\mu }b_{i,\mu }^{\dagger }b_{i,\mu }=1$. The
model Hamiltonian is then rewritten as 
\begin{equation}
H=-\sum_{<i,j>}B_{i,j}^{\dagger }B_{i,j}
\end{equation}%
where $B_{i,j}=\left[
b_{j,4}b_{i,1}+b_{j,1}b_{i,4}-b_{j,3}b_{i,2}-b_{j,2}b_{i,3}\right] $, an
SU(4) symmetric singlet pairing operator.

When a VB order parameter is assumed $\Delta =-\langle B_{i,j}\rangle $ and
the local constraint is imposed by a Lagrangian multiplier $\lambda $, in
terms of a Nambu spinor $\Psi ^{\dagger }(\mathbf{k})=(b_{\mathbf{k}%
,1}^{\dagger },b_{\mathbf{k},2}^{\dagger },b_{\mathbf{k},3}^{\dagger },b_{%
\mathbf{k},4}^{\dagger };b_{-\mathbf{k},1},b_{-\mathbf{k},2},b_{-\mathbf{k}%
,3},b_{-\mathbf{k},4})$, the MF Hamiltonian matrix is obtained as $H_{mf}(%
\mathbf{k})=\lambda +2Z\Delta \gamma _{\mathbf{k}}\Omega _{1}$, where $%
\mathbf{\Omega }_{1}=\sigma _{x}\otimes \sigma _{y}\otimes \sigma _{y}$, and 
$\gamma _{\mathbf{k}}=\frac{1}{Z}\sum_{\mathbf{\delta }}e^{i\mathbf{\ k\cdot
\delta }}$. From the Lagrangian, the bosonic Matsubara Green function is
given by 
\begin{equation}
\mathbf{G}(\mathbf{k,}i\omega _{n})\mathbf{=}\frac{-i\omega _{n}\Omega
_{2}-\lambda +2Z\Delta \gamma _{\mathbf{k}}\Omega _{1}}{\omega _{n}^{2}+ %
\left[ \lambda ^{2}-(2Z\Delta \gamma _{\mathbf{k}})^{2}\right] },
\end{equation}
where $\Omega _{2}=$ $\sigma _{z}\otimes \sigma _{0}\otimes \sigma _{0}$ and
the quasiparticle excitation dispersion $\omega _{\mathbf{k}}\mathbf{=}\sqrt{%
\lambda ^{2}-(2Z\Delta \gamma _{\mathbf{k}})^{2}}$. From the free energy,
the saddle point equations are derived 
\begin{eqnarray}
&&\frac{1}{N}\sum_{\mathbf{k}}\frac{2\lambda }{\omega _{\mathbf{k}}}\left[
2n_{B}(\omega _{\mathbf{k}})+1\right] =3,  \nonumber \\
&&\frac{1}{N}\sum_{\mathbf{k}}\frac{(2Z\gamma _{\mathbf{k}})^{2}}{\omega _{%
\mathbf{k}}}\left[ 2n_{B}(\omega _{\mathbf{k}})+1\right] =Z.
\end{eqnarray}

The dynamic correlation functions of the spin, orbital, and spin-orbital
tensor operators can be calculated by using $S_{i}^{\alpha }=\frac{1}{4}\Psi
^{\dagger }(\mathbf{r}_{i})\Omega _{S}^{\alpha }\Psi (\mathbf{r}_{i})$, $%
T_{i}^{\alpha }=\frac{1}{4}\Psi ^{\dagger }(\mathbf{r}_{i})\Omega
_{T}^{\alpha }\Psi (\mathbf{r}_{i})$, $L_{i}^{\alpha \beta }=\frac{1}{4}\Psi
^{\dagger }(\mathbf{r}_{i})\Omega _{L}^{\alpha \beta }\Psi (\mathbf{r}_{i})$%
, where $\Omega _{S}^{x}=\sigma _{0}\otimes \sigma _{0}\otimes \sigma _{x}$, 
$\Omega _{S}^{y}=\sigma _{z}\otimes \sigma _{0}\otimes \sigma _{y}$, $\Omega
_{S}^{z}=\sigma _{0}\otimes \sigma _{0}\otimes \sigma _{z}$; $\Omega
_{T}^{x}=\sigma _{0}\otimes \sigma _{x}\otimes \sigma _{0}$, $\Omega
_{T}^{y}=\sigma _{z}\otimes \sigma _{y}\otimes \sigma _{0}$, $\Omega
_{T}^{z}=\sigma _{0}\otimes \sigma _{z}\otimes \sigma _{0}$, and $\Omega
_{L}^{\alpha \beta }=$ $\Omega _{S}^{\alpha }\Omega _{T}^{\beta }$. All
fifteen correlation functions associated with $S_{i}^{\alpha }$, $%
T_{i}^{\alpha }$, and $L_{i}^{\alpha \beta }=2S_{i}^{\alpha }T_{i}^{\beta }$
are found to satisfy 
\begin{equation}
\chi _{S}^{\alpha }(\mathbf{q},i\omega _{n})=\chi _{T}^{\alpha } (\mathbf{q}%
,i\omega _{n})=\chi _{L}^{\alpha ,\beta }(\mathbf{q+Q},i\omega _{n}),
\end{equation}
independent of the indices $\alpha $ and $\beta $, implying the the bosonic
MF state is SU(4) symmetric!

On a 2D square lattice and at $T=0$, the conversion from the summations over
momenta to the integrals will be \textit{invalid} as $\lambda \rightarrow
8\Delta $. By separating the divergent terms at $\mathbf{k}^{\ast }\mathbf{=0%
}$, $\mathbf{Q}$ from the summations, the superfluid density $\rho
=4n_{B}(\omega _{\mathbf{k}^{\ast }})/\sqrt{1-(8\Delta /\lambda )^{2}}$ is
defined and obtain $\rho \simeq 0.107$, $\Delta \simeq 1.3159$, and $\lambda
\simeq 10.5271$. The dynamic correlation function is evaluated, leading to
the imaginary part of the dynamic susceptibility for $\omega >0$, 
\begin{eqnarray*}
&&\func{Im}\chi _{S}(\mathbf{q},\omega )\approx \frac{\pi }{8}\int \frac{%
d^{2}\mathbf{k}}{(2\pi )^{2}}\left[ 1-\frac{\lambda ^{2}-(8\Delta
)^{2}\gamma _{\mathbf{k}}\gamma _{\mathbf{k+q}}}{\epsilon _{\mathbf{k}%
}\epsilon _{\mathbf{k+q}}}\right] \\
&&\hspace{0.5cm}\text{ \ \ \ }\times \left[ n_{B}(\epsilon _{\mathbf{k+q}%
})+n_{B}(\epsilon _{\mathbf{k}}\mathbf{)+}1\right] \delta \left( \omega
-\epsilon _{\mathbf{k}}\mathbf{-}\epsilon _{\mathbf{k+q}}\right) .
\end{eqnarray*}%
As $\mathbf{q}\rightarrow \mathbf{Q}$ with $\lambda \rightarrow 8\Delta $, $%
\func{Im}\chi _{S}(\mathbf{Q},\omega )$ is divergent. From the fluctuation
dissipation theorem, the dynamic structure factor is obtained $S(\mathbf{Q}%
,\omega )\approx N\left( \rho /\sqrt{2}\right) ^{2}2\pi \delta (\omega )$,
yielding to a long-range AF ordered state with the staggered magnetizations $%
m_{S}\approx 0.07566$.

These results \emph{disagree} with the prediction of a spin liquid state
with a finite gap by the quantum Monte Carlo calculations on a $12\times 12$
square lattice \cite{santoro}, and is believed due to their small system
sizes \cite{zhang-shen}. Interestingly, the latest quantum Monte Carlo
simulations on much larger lattice sizes \cite{harada} have showed that the
convergence of staggered magnetization to a finite value can be seen for
systems larger than $32\times 32$ lattices, and the estimated staggered
magnetization is $m_{S}\sim 0.0703$, which is nearly the same value as ours.

The author would like to thank S. Q. Shen for his useful discussion.


\begin{thebibliography}{9}
\bibitem{santoro} G. Santoro, \textit{et al.}, Phys. Rev. Lett. \textbf{83},
3065 (1999).

\bibitem{aa1988} D. P. Arovas and A. Auerbach, Phys. Rev. B \textbf{38}, 316
(1988).

\bibitem{zhang-shen} G.-M. Zhang and S. Q. Shen, Phys. Rev. Lett. \textbf{87}%
, 157201 (2001).

\bibitem{martins} M. J. Martins and B. Nienhuis, Phys. Rev. Lett. \textbf{85}%
, 4956 (2000).

\bibitem{harada} K. Harada, N. Kawashima, and M. Troyer, Phys. Rev. Lett. 
\textbf{90}, 117203 (2003).
\end{thebibliography}
\end{document}